\documentclass[12pt]{iopart}
\newcommand{\beq}{\begin{equation}}
\newcommand{\eeq}{\end{equation}}
\newcommand{\beqa}{\begin{eqnarray}}
\newcommand{\eeqa}{\end{eqnarray}}
\newcommand{\ba}{\begin{array}}
\newcommand{\ea}{\end{array}}

\begin{document}

\title[Kirzhnits gradient expansion for a D-dimensional Fermi gas]
{Kirzhnits gradient expansion for a D-dimensional Fermi gas} 

\author{Luca Salasnich} 

\address{CNR-INFM and CNISM, Unit\`a di Padova, 
Dipartimento di Fisica ``Galileo Galilei'', Universit\`a
di Padova, Via Marzolo 8, 35131 Padova, Italy} 

\ead{salasnich@pd.infn.it}

\begin{abstract} 
For an ideal D-dimensional Fermi gas under generic external confinement 
we derive the correcting coefficient $(D-2)/3D$ 
of the von Weizsacker term in the kinetic energy density. 
To obtain this coefficient we use the Kirzhnits 
semiclassical expansion of the number operator up to the 
second order in the Planck constant $\hbar$. 
Within this simple and direct approach we determine 
the differential equation of the 
density profile and the density functional of the Fermi gas. 
In the case $D=2$ we find that the Kirzhnits 
gradient corrections vanish to all order in $\hbar$. 
\end{abstract} 

\maketitle

\section{Introduction} 

At ultra-low temperatures macroscopic effects of 
quantum statistics are observable 
with bosonic vapors of neutral alkali-metal atoms 
\cite{book-pethick,book-stringari}. 
In the last years quantum degeneracy 
has been achieved also with fermionic atoms 
\cite{demarco,truscott,modugno,greiner,joachim}. 
For fermions, it has been predicted that a reduced dimensionality  
strongly modifies density profiles \cite{schneider,sala0,sala1,vignolo}, 
collective modes \cite{minguzzi} and stability of mixtures 
\cite{das,sala2}. In all these papers a Thomas-Fermi-von Weizs\"acker 
density functional approach is used \cite{thomas-fermi,von}.  
\par 
Some years ago, Holas, Kozlowski and March \cite{march} 
derived for a D-dimensional ideal Fermi gas a correcting coefficient 
to the von Weizsacker term in the kinetic energy density. 
They used a response-function approach based on the knowledge
of the static susceptibility in a D-dimensional space
\cite{holas}. The aim of this paper is to recover the result 
of \cite{march} by using the Kirzhnits 
semiclassical expansion of the number operator \cite{kirzhnits}. 
We shall show that the Kirzhnits expansion in powers of $\hbar$ 
is a direct and simple approach to get the density profile 
and the energy of a D-dimensional ideal Fermi gas under generic 
external confinement. 

\section{Thomas-Fermi-von Weizsacker formulation} 

For an ideal D-dimensional spin-polarized Fermi gas 
in a hypercubic box of side $L$ 
the number of particles at zero temperature is given by 
\beq 
N = \int {d^D{\bf r}\, d^D{\bf p}\over (2\pi \hbar)^D} \;  
\Theta\left({\bar{\mu}} - {p^2\over 2m} \right) \; ,  
\label{total-number}
\eeq 
where $\Theta(x)$ is the Heaviside step function. 
The chemical potential $\bar{\mu}$, 
that is the Fermi energy, 
fixes the number of particles $N$. 
In the formula we have substituted 
the discrete summation over box quantum numbers with the 
2D-dimensional integral over single-particle phase-space variables 
$({\bf r},{\bf p})$. This approximation becomes exact in the limit 
$L,N\to \infty$ and $n=N/L^D$ constant. 

One can easily integrate over ${\bf r}$ and ${\bf p}$ 
in Eq. (\ref{total-number}) and finds 
the following expession for the uniform density 
\beq 
n = \left({m\over 2\pi\hbar^2}\right)^{D/2} 
{D\over 2 \Gamma({D\over 2}+1)} F_D(\bar{\mu} ) \; , 
\label{n-uniform} 
\eeq 
where
\beq
F_D(\bar{\mu}) = \int_0^{\infty} dy \, y^{D/2-1} \, 
\Theta(\bar{\mu} - y) = {2\over D} \bar{\mu}^{D/2} \; ,  
\eeq 
measures the number of states 
with energy less or equal to $\mu$. 

In many papers the inclusion of an external potential $U({\bf r})$ is 
treated within the so-called local density approximation, 
i.e. by simply using a local chemical potential 
\beq 
\mu({\bf r}) = \bar{\mu} - U({\bf r}) \;   
\label{local-chemical}
\eeq 
in the Eq. (\ref{n-uniform}), which becomes 
\beq
n({\bf r}) = \left({m\, \mu({\bf r})\over 2\pi \hbar^2}\right)^{D/2} 
{1 \over \Gamma({D\over 2}+1)} \; . 
\label{n-local-basic}
\eeq 
This expression of the local density $n({\bf r})$ can be obviously 
obtained from Eq. (\ref{total-number}) by inserting on it 
Eq. (\ref{local-chemical}) and integrating over momenta 
${\bf p}$. Eq. (\ref{n-local-basic}), also called Thomas-Fermi 
formula of the density profile\cite{thomas-fermi}, can be written 
in the following form 
\beq 
{2\pi\hbar^2\over m} \Gamma\left({D\over 2}+1\right)^{2/D} \, 
n({\bf r})^{2/D} + U({\bf r}) =\, \bar{\mu} \; , 
\label{explicit} 
\eeq
to make explicit the dependence of the chemical 
potential $\bar{\mu}$ on the local density $n({\bf r})$. 
It is important to stress that Eq. (\ref{n-local-basic}), and also 
Eq. (\ref{explicit}), 
is semiclassical because it is derived by neglecting 
the fact that in quantum mechanics 
the kinetic operator ${\hat T} =-\hbar^2\nabla^2/(2m)$ and 
the operator ${\hat \mu}=\mu({\bf r})=\bar{\mu} - U({\bf r})$ 
do not commute. Historically, it was von Weizs\"acker \cite{von} 
the first to introduce a gradient correction (in the case $D=3$) 
which takes into account phenomenologically the increase of 
kinetic energy due spatial variation of the density. 
Including the gradient correction of von Wiezs\"acker in 
Eq. (\ref{explicit}) one has 
\beq 
{\hbar^2\over 2m}
\left[ {1\over 4} 
{(\nabla n({\bf r}))^2\over n({\bf r})^2} - 
{1\over 2} {\nabla^2 n({\bf r})\over n({\bf r})} \right] 
+ {2\pi \hbar^2\over m} \Gamma\left({D\over 2}+1\right)^{2/D} 
\, n({\bf r})^{2/D} + U({\bf r}) = \, \bar{\mu} \; . 
\label{weizs}
\eeq
Remarkably, the gradient correction of von Weizs\"acker does not 
depend on the dimensionality of the system, as shown in \cite{march}.  

\section{Kirzhnits gradient expansion} 

Let us now consider a full quantum-mechanical treatment. 
In quantum mechanics the number operator is 
\beq 
{\hat N} = \Theta\left( {\hat \mu} - {\hat T} \right) \; .   
\eeq 
The exact local density $n({\bf r})$ can be written by using 
the number operator ${\hat N}$, 
the eigenstate $|{\bf r}\rangle$ 
of the position operator, and the eigenstate 
$|{\bf p}\rangle$ of the momentum operator 
in the following way
\beq
n({\bf r}) = \langle {\bf r}| {\hat N} |{\bf r} \rangle
= \int {d^D{\bf p}\over (2\pi \hbar)^{D/2}}
\langle{\bf r}|{\hat N} |{\bf p}\rangle e^{-i{\bf p}
\cdot {\bf r}/\hbar} \; , 
\label{n-exact} 
\eeq 
where the last expression has been obtained by inserting the 
completeness 
\beq
\int d^D{\bf p}\, 
|{\bf p} \rangle \langle {\bf p}| = 1 \; ,
\eeq
and the formula $\langle {\bf p} |{\bf r} \rangle =
e^{-i{\bf p}\cdot {\bf r}/\hbar}/(2\pi\hbar)^{D/2}$. 
According to Kirzhnits \cite{kirzhnits,dreizler}, 
one can expand the operator ${\hat N}$ 
at the second order of $\hbar$ in the following way 
$$
{\hat N} |{\bf p} \rangle = 
\Theta\left( {\hat \mu} - {\hat T} \right)
|{\bf p} \rangle = \Theta({\hat \mu} - T) | {\bf p}\rangle 
+ {1\over 2} \Theta''({\hat \mu}-T) [{\hat \mu},{\hat T}] |{\bf p} \rangle 
$$
\beq 
- {1\over 6} \Theta'''({\hat \mu}-T) 
\left\{ [{\hat \mu}, [{\hat \mu},{\hat T}]] 
+ [{\hat T},[{\hat \mu},{\hat T}]] \right\} |{\bf p} \rangle 
+ {1\over 8} \Theta''''({\hat \mu}-T) [{\hat \mu},{\hat T}]^2 
|{\bf p} \rangle 
\label{kirzh} 
\eeq
where $[{\hat \mu},{\hat T}]={\hat \mu}{\hat T}-{\hat T}{\hat \mu}$ 
and ${\hat T}|{\bf p} \rangle = p^2/(2m) |{\bf p} \rangle 
= T |{\bf p} \rangle$. 
Here $\Theta'(x)$ is the first derivative of $\Theta(x)$ with respect 
to $x$, i.e. $\Theta'(x)=\delta(x)$ is the Dirac delta function, 
$\Theta''(x)$ is the second derivative of $\Theta(x)$, and so on. 
It is important to observe that Eq. (\ref{kirzh}) does not depend 
on the dimensionality of the system. Kirzhnits used 
Eq. (\ref{kirzh}) in the special case $D=3$ \cite{kirzhnits}, 
while here we consider it for a generic $D$. 
Note that higher order terms of the Kirzhnits expansion 
of the number operator ${\hat N}$ 
involve higher derivatives of the Heaviside function 
$\Theta({\hat \mu} - T)$ \cite{kirzhnits,dreizler}. 
\par
The term $\langle {\bf r}|{\hat N}|{\bf p}\rangle$ 
in Eq. (\ref{n-exact}) can be now calculated 
by using the Kirzhnits expansion. 
In this way, after straightforward but lengthy 
manipulations, at the second order in $\hbar$ 
Eq. (\ref{n-exact}) becomes 
\beqa
n({\bf r}) = \left({m\over 2\pi \hbar^2}\right)^{D/2}  
{D \over 2 \Gamma({D\over 2}+1)} 
\Big\{ F_D(\mu({\bf r})) 
+ {\hbar^2\over 2m} \Big[ {1\over 6} \nabla^2 \mu({\bf r}) 
F_D''(\mu({\bf r})) 
\nonumber 
\\
+ {1\over 12} 
\left(\nabla \mu({\bf r})\right)^2 F_D'''(\mu({\bf r})) 
\Big] \Big\} \; , 
\label{figata}
\eeqa
where $F_D'(\mu )$ is the first derivative of $F_D(\mu )$ 
with respect to $\mu$, $F_D''(\mu )$ is the second 
derivative, and so on. 
In the Kirzhnits expansion, after integration 
over momenta, each term containing a $n$-th derivative 
of the Heaviside function $\Theta({\hat \mu} - T)$ gives rise 
to a term containing a $(n-1)$-th derivative 
of the function $F_D(\mu)$. As a consequence, in the special 
case $D=2$, where $F_2(\mu)=\mu$, only the leading term 
of the Kirzhnits expansion is different from zero, i.e. 
with $D=2$ the Kirzhnits gradient corrections vanish 
to all order in $\hbar$. 

In the case $D=3$, i.e. the three-dimensional Fermi gas, 
Eq. (\ref{figata}) becomes  
\beq 
n({\bf r}) = \left({2m\, \mu({\bf r})\over \hbar^2}\right)^{3/2}  
{1\over 6\pi^2} \Big\{ 1 + {\hbar^2\over 2m} 
\Big[ {1\over 8} {\nabla^2 \mu({\bf r}) \over \mu({\bf r})^2} - {1\over 32} 
{\left(\nabla \mu({\bf r})\right)^2\over \mu({\bf r})^3} 
\Big] \Big\} \; ,  
\eeq 
that is exactly the formula reported in the book 
of Dreizler and Gross \cite{dreizler}, which considers only 
the three-dimensional case. In the case $D=2$, i.e. 
a two-dimensional Fermi gas, we have seen that 
all the quantum corrections are zero, 
while in the $D=1$ case these corrections are in general 
finite, and in particular the $\hbar^2$ ones. 

We observe that the vanishing of quantum corrections 
in the Kirzhnits expansion 
for the two-dimensional ideal Fermi gas is a consequence 
of the linear number of states, i.e. constant density of states, 
in this specific dimension. 
Obviously, this does not mean that with $D=2$ the Thomas-Fermi 
formula of the density profile is exact; it simply means that  
the Kirzhnits expansion fails in the $D=2$ case. 
For instance, recently Brack and van Zyl \cite{zyl} 
have found that, for the a 2D ideal gas of fermions 
under hamonic confinement, the quantum kinetic energy 
is exactly reproduced by the Thomas-Fermi kinetic energy 
without gradient corrections, but only using 
the exact density of states, which is different from 
the Thomas-Fermi one \cite{zyl}. 

Eq. (\ref{figata}) has a serious turning point problem: 
The local chemical potential $\mu({\bf r})=\bar{\mu}-U({\bf r})$ 
becomes zero at the classical turning points, 
where $\bar{\mu}=U({\bf r})$. 
Therefore, the gradient terms in Eq. (\ref{figata}) diverge 
at the turning points. To overcome this problem we invert 
Eq. (\ref{figata}) finding $\mu({\bf r})$ as a function of 
$n({\bf r})$. Obviously, due to the gradients 
this inversion cannot be done 
exactly but only in a perturbative way. First we note 
that Eq. (\ref{figata}) has the following structure 
\beq 
n({\bf r}) = A\, \mu({\bf r})^{D/2} 
\left( 1 + {\hbar^2\over 2m} B[\mu({\bf r}),\nabla \mu({\bf r})] 
\right) \; , 
\eeq 
where $A = (m/(2\pi\hbar^2))^{D/2}/\Gamma(D/2+1)$. Then 
from this expression we get 
\beqa
\mu({\bf r}) = { A^{-2/D} n({\bf r})^{2/D} \over 
\left( 1 + {\hbar^2\over 2m} B[\mu, \nabla \mu({\bf r})] \right)^{2/D} } 
\nonumber 
\\
= A^{-2/D} n({\bf r})^{2/D} \left\{ 1 - 
{2\over D} {\hbar^2\over 2m} B[\mu({\bf r}), \nabla \mu({\bf r})] 
+ O(\hbar^4) \right\} \; ,  
\eeqa 
where $O(\hbar^4)$ contains terms of order $\hbar^4$. Finally, 
we find, at the second order in $\hbar$, this expression 
\beqa
{\hbar^2\over 2m} \left({D-2\over 3D}\right) 
\left[ {1\over 4}
{(\nabla n({\bf r}))^2\over n({\bf r})^2} -
{1\over 2} {\nabla^2 n({\bf r})\over n({\bf r})} \right] 
\nonumber 
\\
+ {2\pi\hbar^2\over m} \Gamma\left({D\over 2}+1\right)^{2/D} 
\, n({\bf r})^{2/D} + U({\bf r}) = \, \bar{\mu} \; .
\label{final}
\eeqa
This result differs from that of von Weizs\"acker, 
Eq. (\ref{weizs}), 
simply because there is a the factor $(D-2)/3D$ 
in the gradient terms. 
We stress that the von Weizs\"acker gradient terms are purely 
phenomenological, while the Kirzhnits gradient terms 
are based on a perturbative expansion of the exact number 
operator. In the scientific literature often the 
von Weizs\"acker gradient terms 
are modified by inserting a coefficient $c$ 
in front of them. For instance, with $D=3$, 
March and Tosi \cite{tosi} suggest $c=1/3$, 
while Zaremba and Tso \cite{zaremba} suggest $c=1/36$. 
Remarkably, both the phenomenological improvements 
disagree with the Kirzhnits result $c=1/9$. 

As previously discussed, the coefficient $(D-2)/3D$ was 
found some years ago by Holas, Kozlowski and March \cite{march} 
by using a response function approach based on the knowledge 
of the static susceptibility in a D-dimensional space 
\cite{holas}. Here we have instead applied 
the more direct Kirzhnits expansion. For $D=3$ the factor gives 
the familiar coefficient $1/9$, for $D=2$ the factor is zero, 
and for $D=1$ the factor is $-1/3$. 

It is straightforward to show that the differential 
equation (\ref{final}) is obtained by minimizing the energy 
density functional 
\beqa 
E[n({\bf r})] = 
\int d^D{\bf r} \, \Big\{ 
\left({D-2\over 3D}\right){\hbar^2 \over 8 m} 
{(\nabla n({\bf r}))^2\over n({\bf r})} 
\nonumber 
\\
+ {2\pi\hbar^2\over m}{D\over D+2} 
\Gamma\left({D\over 2}+1\right)^{2/D} n({\bf r})^{(D+2)/D} + 
U({\bf r})\, n({\bf r}) \Big\} 
\label{ppp}
\eeqa
with the constraint 
\beq 
N = \int d^D{\bf r} \, n({\bf r}) \; , 
\eeq 
which fixes the chemical potential $\bar{\mu}$. Eq. (\ref{ppp}) 
can be used as a starting point to investigate mixtures of 
interacting Fermi atoms and also electrons in nanotubes. 

\section{Conclusions} 

We have been able to determine the equations 
for the energy and the density profile of a D-dimensional ideal Fermi 
gas with external confinement by using the Kirzhnits expansion. 
These equations, which are correct 
at the second order in the Planck constant $\hbar$, 
can be improved by including higher order terms of the 
Kirzhnits expansion of the number operator \cite{kirzhnits}. 
In fact, in some cases, $\hbar$-expansions can be calculated 
to all orders and resummed to give the exact result \cite{sala-wkb}. 
In the special case $D=2$ we have found, by using the Kirzhnits expansion, 
that all $\hbar$ corrections to the Thomas-Fermi formula 
of the density profile are zero. This is a direct consequence 
of the constant density of states in a two-dimensional ideal gas 
and it means that the Kirzhnits expansion is not reliable under 
these conditions.  
We remind that gradient corrections to 
the Thomas-Fermi formula of the density profile have been analyzed 
with different semiclassical techniques in previous studies 
\cite{march,brack}. On the other hand, to our knowledge, 
this is the first time 
that the leading terms of the Kirzhnits expansion
have been directly obtained for an ideal Fermi gas in a D-dimensional space
and generic confining potential. 

\section*{Acknowledgments}

This work has been partially supported by GNFN-INdAM and 
by Fondazione CARIPARO. The author thanks Prof. Flavio Toigo 
for many enlightening discussions and suggestions.

\end{document}